# A Modified IEEE 802.15.6 MAC Scheme to Enhance Performance of Wireless Body Area Networks in E-health Applications

Md. Abubakar Siddik[1], Most. AnjuAra Hasi[1], JakiaAkter Nitu[1],
Sumonto Sarker[1], Nasrin Sultana[1] and Emarn Ali[2]

[1]Department of Electronics and Communication Engineering, Hajee Mohammad Danesh
Science and Technology University, Dinajpur, Bangladesh
[2]Department of Computer Science and Engineering, Hajee Mohammad Danesh Science
and Technology University, Dinajpur, Bangladesh

## Abstract

*The recently released IEEE 802.15.6 standard specifies several physical (PHY) layers and medium access control (MAC) layer protocols for variety of medical and non-medical applications of Wireless Body Area Networks (WBAN). Themedical applications of WBAN hasseveral obligatory requirements and constrains viz. high reliability, strict delaydeadlines and low power consumption.The standard IEEE 802.15.6 MAC scheme is not able to fulfil the all requirements of medical applications of WBAN. To address this issue we propose anIEEE 802.15.6-based MAC schemethat is the modification of superframe structure, user priorities and access mechanism of standard IEEE 802.15.6 MAC scheme. The proposed superframe has three access phases: random access phases (RAP), manage access phases (MAP) and contention access phase (CAP). The proposed four user priorities nodes access the channel during RAP using CAMA/CA mechanism with a large value of contention window. The proposed MAC scheme uses RTS/CTS access mechanism instead of basic access mechanism to mitigate the effect of hidden and expose terminal problem. Moreover, we develop an analytical model to evaluate the performance of proposed MAC scheme and solve the analytical model using Maple. The results show that the modified IEEE802.15.6 MAC scheme achieve the better performance in terms of reliability, throughput, average access delay, energy consumption, channel utilization and fairness compared to standard IEEE 802.15.6 MAC scheme in E-health applications.*

## Keywords

## 1. Introduction

In the era of advanced technology, increasing needs of aging population, rising costs of healthcare, limited healthcare resources, especially during worldwide pandemic like COVID-19, have triggered the concepts of Wireless Body Area Networks (WBANs) as a primary part of the ubiquitous Internet of Medical Things (IoMT) systems and received considerable attention in the academy and industry. The WBAN is composed of a limited number of tiny, low-power, low-cost, wearable or implantable, intelligent, and heterogeneous medical sensors that are deployed in, on or around the proximity of the human body for continuously sensing vital physiological signals. The sensed signals are then aggregated at a coordinator via a short-range, low power wireless communication, provided by IEEE 802.15.6 standard [1] and forwarded to the servers for further analysis. The WBAN offers numerous medical and non-medical applications in ubiquitous healthcare, military and defence, sports and fitness, and entertainment fields for improving the quality of human life, described in detail in [2]–[4]. Each WBAN application hassome specific QoS requirements like reliability, latency, security, and power consumption [5].

The communication architecture of WBANs is the combination of three different tiers: Tire-1, known as intra-WBAN communication, which comprises the communication between node andcoordinator. Tire-2, known as inter-WBAN communication, which involves the communication between coordinator and one or more access point (AP). Finally, Tire-3, known as beyond-WBAN communication, acts as a bridge between the tire-2 and external network [5]–[6]. Earlier, several communication standards were used as an access network standard in WBAN, such as IEEE 802.11 (WiFi), IEEE 802.15.1 (Bluetooth) and IEEE 802.15.4 (Zigbee). However, these standards are not able to fulfil the requirements of WBAN applications [7].The IEEE 802.15 Task Group 6 has developed a new communication standard, IEEE 802.15.6 that specifies the physical (PHY) and medium access control (MAC) layers specifications for the WBAN.

The protocol ofMAC layer, lower sub-layer of the data link layer of the OSI model, is responsible for sharing channel among the contending nodes such that channel is distributed either fairly or priority basis with conforming less frame collision.The MAC protocol also defines several parameters like the number of control packets, size of overhead, collision probability, superframe structure, data priority whichare used to control the overall performance of the network.Thestandard IEEE 802.15.6 MAC scheme is not able to fulfill the all requirements of WBAN because it has some unique characteristics and requirements. So a new MAC protocol is necessary to fulfill the requirements of WBAN applications. In this paper, we propose a MAC scheme that is the modification of standard IEEE 802.15.6 MAC scheme. The proposed MAC scheme will enhance the overall performance of the WBAN in terms of reliability, throughput, average access delay, energy consumption, channel utilization and fairness.

The rest of the paper is organized as follows. Section 2 addresses the related works and Section 3 brief overviews the standard IEEE 802.15.6 MAC scheme. A modified IEEE802.11 MAC scheme is proposed in Section 4.In Section 5 a Markov chain-based analytical model for modified IEEE 802.15.6 MAC scheme is designed. The expressions for different performance metrics is derived in Section 6. The performance metrics evaluation of this work are presented in Section 7. Finally, Section 8 concludes the paper and gives the future work outlines.

## 2. RELATED WORKS

There are several literatures [8]–[16] that explore the performance analysis of the IEEE 802.15.6 standard and investigate the effects of different prioritizing parameters on the performance metrics. In [8], Sarkar et al. proposed a user priority wise and Markov chain-based analytical model of IEEE 802.15.6 CSMA/CA for noisy channel and situation traffic condition to evaluate some performance metrics viz. reliability, throughput, energy consumption and average delay by taking into account the acknowledgement time or timeout after transmitting packets. The effects of channel condition, packet size and data rate on the performance metrics are also presented. The major finding of this paper is that the instead of eight different UPs, five would have been sufficient.In [9], the authors proposed an analytical model for all UPs to investigate the impact of access phase lengths on throughput and average backoff time. The conclusion of this study is thatEAP1, EAP2 and RAP2 are unnecessary and poor medium utilization is achieved under high traffic load.Jacob et al. [10] developed a sleep mechanism to extend the network lifetime and designed a user priority wise and Markov chain-based analytical model to measure the performance of IEEE 802.15.6 CSMA/CA for ideal channel environment and non-saturation traffic condition.The author summarizedthe three major findings: (i) increasing length of EAP serves to increase the average delay of lower user priority nodes (ii) highest user priority ($UP_7$) emergency nodes generate packet occasionally, which in turn justifies a short EAP fraction compared with RAP,and (iii) sleep mechanism increase the lifetimes of sensors.Khan et al. [11] evaluated normalized throughput and mean frame service time of CSMA/CA mechanism by developing a user priority wise and Markov chain-based analytical model and assuming that the channel is ideal and nodes contain non-saturated traffic. The major observations of the authors are that low user priorities utilize the medium poorly and face higher delay,and use of different access phases degrades the overall network performance.Yuan et al. [12] analyzed the performance of IEEE 802.15.6-based WBAN in presence of intra-WBAN and inter-WBAN interference by designing a user priority wise and Markov chain-based analytical model for ideal channel and saturation traffic condition of nodes. The authors concluded that 70% throughput decreases and 50% delay increases due to interference generated by neighbor WBANs.In [13]–[14], the authors designed a Markov chain-based complete analytical model of IEEE 802.15.6 CSMA/CA by considering most of the constrains of IEEE 802.15.6 standard and evaluated the performance parameters viz. normalized throughput, successful transmission probability, mean waiting time and average time between two successive access.The authors concluded some major findings: (i) EAP is not necessary, (ii) four user priorities are sufficient (iii) short EAP and RAP lead to poor utilization of available bandwidth, (iv) medium is unfairly accessed by highest user priority, and (v) other user priorities ($UP_0$ - $UP_6$) except highest user priority are starving.In [15]–[16], performance of IEEE 802.15.6 CSMA/CA is evaluated using RTS/CTS access mechanism instead of basic access mechanism. The summary of these studies is that the RTS/CTS access mechanism has potentiality to enhance the performance in presence of hidden and exposed terminals.The findings of the above literatures motivate us to design a new MAC protocol for improving the performance with maintaining the desired QoSby modifying the number of user priorities, superframe structure and access mechanism of the IEEE 802.15.6 standard.

There are several MAC protocols based on IEEE 802.15.6 standard have been proposed for WBAN by modifying the mechanisms and/orparameters ofthe IEEE 802.15.6 standard [18-20]. Yuan et al. [19] proposed an adaptive MAC (A-MAC) protocol that defines three user priorities node according to the type of service andimproves the superframe structure based on IEEE 802.15.6which consists of four adjustable access phases according to the traffic flow: beacon phase, contention access phase, non-contention access phase and inactive phase.The contention access phase is further divided into three adjustable sub-phases according to the data priority that are accessed bythe three user prioritiesindividually.The simulation results show that the proposed A-MAC protocol is achieved better performance compared to CA-MAC protocol and IEEE 802.15.6 MAC protocol in terms of normalized throughput, average energy consumption and average delay.However, in the contention access phase, lower user priority nodes suffer from a bounded delay due to presence of sub-phases.Moreover, the three user priorities are not sufficient to cover all traffic instead of eight user priorities defined in IEEE 802.15.6 standard.Sultana et al.[20] proposed a collaborative medium access control (CMAC) protocol for critical and noncritical traffic whichare generated in wireless body area network.The protocol classify the critical traffic into three priorities based on delay deadlines whereas noncritical traffic is divided into three priorities based on energy level. Moreover, the authors design four superframeswhich are used to transmit different critical and noncritical traffic based on event categories.The results show that critical traffic achieve least delay compared to noncritical traffic. However, in this study other vital performance parameters viz. throughput and energy consumption are not evaluated and their donot compare their results with exiting MAC protocols.In [21], the authors developed a superframe that have three access phases: EAP, RAP and MAP, and modified the contention window bound for each user priority. The proposed scheme gains worst latency of medical traffic and improve mean delay time of non-medical traffic compare to IEEE 802.15.6 MAC scheme. However, in this study other vital performance parameters viz. throughput and energy consumption are not consider.

According to the above research background, the IEEE 802.15.6 MAC scheme has some redundant access phases (EAP2, RAP2 and MAP2) in the superframe structure that create complexity to manage the access process. Moreover, the EAP access phasesare only used for the highest user priority ($UP_7$) nodes which generate traffic occasionally, and thus the result is poor medium utilization.The eight user priorities of IEEE 802.15.6 MAC scheme are redundant to cover traffic generated from different realistic medical applications of WBAN.In addition, the basic access mechanism of IEEE 802.15.6 MAC scheme suffer from hidden terminal and exposed terminal problem for inter-WBAN communication.

To address above issue, we propose a modified IEEE 802.15.6 MAC scheme that is the modification of number of user priorities, access mechanism and superframe structure of standard IEEE 802.15.6 MAC scheme. The proposed MAC scheme improve the performance in terms of reliability, throughput, average access delay, energy consumption, channel utilization and fairness.

## 3. STANDARD IEEE 802.15.6 MAC SCHEME

The IEEE 802.15.6 standard [1] introduces eight user priorities($UP_i$where$i \epsilon [0,7]$)which are differentiated by the values of the minimum contention window ($CW_{i,min}$) and maximum contention window ($CW_{i,max}$). It defines three access modes: beacon mode with superframes, non-beacon mode with superframes, and non-beacon mode without superframes. In beacon mode with superframes, the coordinator periodically sends a beacon at the beginning of every superframe in the network and the nodes are synchronized by receiving it.The superframe is divided into seven access phases, i.e., Exclusive Access Phase 1 (EAP1), Random Access Phase 1 (RAP1), Managed Access Phase 1 (MAP1), Exclusive Access Phase 2 (EAP2), Random Access Phase 2 (RAP2), Managed Access Phase 2 (MAP2), and Contention Access Phase (CAP), shown in Figure 1.The EAP1 and EAP2 are used for $UP_7$, and the RAP1, RAP2 and CAP are used for all UPs.The default access mechanism of standard IEEE 802.15.6 MAC scheme is the basic access mechanism where data and acknowledge frame are exchanged to complete a transmission.As a result, the duration of successful transmission and collision transmission need same amount of time. The frame format of data and acknowledge frame and success and collisiontime of basic access mechanism are shown in Figure 2.

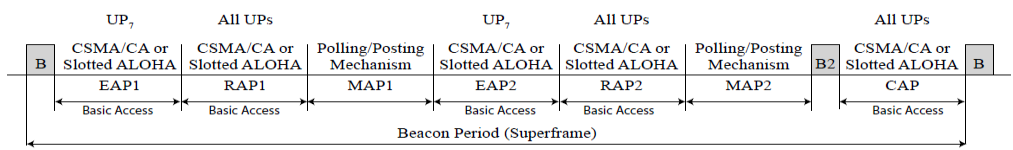

Figure 1. Superframe of standard IEEE 802.15.6 MAC scheme

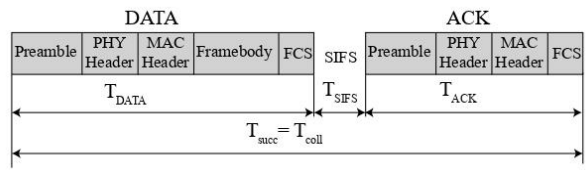

Figure 2. Success time and collision time of basic access mechanism for standard IEEE 802.15.6 MAC scheme

In order to access the channel, the standard 802.15.6 MAC scheme defines three access mechanisms:random access mechanism, improvised and unscheduled access mechanism and scheduled access mechanism.The random access mechanism (either CSMA/CA or slotted

ALOHA) is used in EAP, RAP and CAP access phases. The access procedure of CSMA/CA mechanism is given below:

According to CSMA/CA mechanism of IEEE 802.15.6 standard, a $UP_i$ node shall maintain a backoffcounter and contention window ($W_{i,j}$) to get access of the channel. The node shall set its backoff counter value to a randomly chosen integer over $[1, W_{i,j}]$ and decrease its backoff counter by one for each idle CSMA slot. The node will transmit one frame over the channel if backoff counter reaches zero. The contention window selection, locking and unlocking process of the backoff counter during the contention period depends on channel and transmission state. A new contention window is selected as

- The $UP_i$ node shall set $W_{i,j}$ to $CW_{i,min}$ if it transmits the frame successfully or it contains a new frame for transmission.
- If the $UP_i$ node does not transmit its frame successfully, it shall keep the $W_{i,j}$ unchanged if this is the $j$th time the node had failed consecutively, where $j$ is an odd number; otherwise, the $W_{i,j}$ is doubled.
- The $UP_i$ node shall set $W_{i,j}$ to $CW_{i,max}$ if the new contention window exceeds $CW_{i,max}$ and the number of attempts of transmission does not exceed the retry limits.

When any of the following events occurs during the contention period, the node shall lock its backoff counter.

- The channel is sensed busy due to another transmissions.
- The current access phase does not permit the $UP_i$ node for the current transmission attempt.
- The current time is at the start of a CSMA slot within the permitted access phase of $UP_i$ node but the time between the end of the CSMA slot and the end of the permitted access phase is not long enough for a frame transmission.

The backoff counter will be unlocked when the channel has been idle for SIFS time within the permitted access phase of $UP_i$ node and the time interval between the current time and the end of the permitted access phase is long enough for a frame transmission. After unlocking the backoff counter, the node shall start the backoff procedure by observing the channel conditions and transmit one frame over the channel if backoff counter reaches zero.

## 4. MODIFIED IEEE 802.15.6 MAC SCHEME

To improve the overall performance the final objectives of this study are to increase the channel utilization and throughput as well as to reduce the frame collision probability and the channel access delay. This would be possible if the superframe structure and user priorities are modified by considering real traffic pattern of E-health applications of WBAN so that the redundancy of access phases and traffic differentiation can be minimized. The collision probability due to hidden terminal also can be reduced by using RTS/CTS access mechanism. In this paper, we propose a MAC scheme for E-health applications of WBAN called modified IEEE 802.15.6 MAC scheme which is the modification of standard IEEE 802.15.6 MAC scheme. We modify traffic differentiation, superframe structure and access mechanism for beacon mode with superframe. We develop a new traffic differentiation where two user priorities of standard IEEE 802.15.6 MAC scheme are combined into one user priority and each new user priority is denoted by the former lower user priority. We also define the size of contention window for new user priorities which are equal to the larger contention window size of the former lower user priority.

Consequently, we get four user priorities for the proposed MAC scheme. The proposed superframe is divided into three access phases, i.e., Random Access Phase (RAP), Managed Access Phase (MAP), and Contention Access Phase (CAP) shown in Figure 3.The RAP and CAP are used for all UPs. We also propose that the RTS/CTS access mechanism is used in RAP and CAP access phases with CSMA/CA access protocol instead of basic access mechanism introduced by IEEE 802.15.6 standard.In RTS/CTS access mechanism, four frames (RTS, CTS, DATA and ACK) are exchanged to complete a transmission. The frame format of these frames and success and collision time of RTS/CTS access mechanism are shown in Figure 4. To access the channel, the propose MAC scheme also use CSMA/CA mechanism that is same as IEEE 802.15.6 CSMA/CA, described in Section 3. The comparison of MAC parameters for the standard IEEE 802.15.6 MAC scheme and the modified IEEE 802.15.6 MAC scheme is summarised in Table 1.

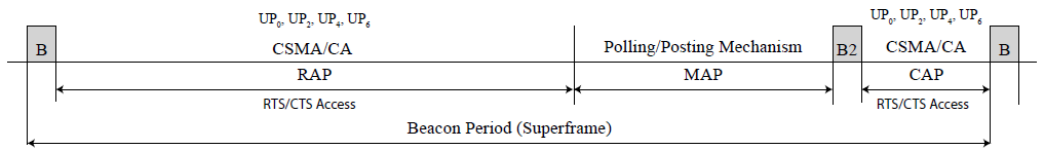

Figure 3. Superframe of modified IEEE 802.15.6 MAC scheme

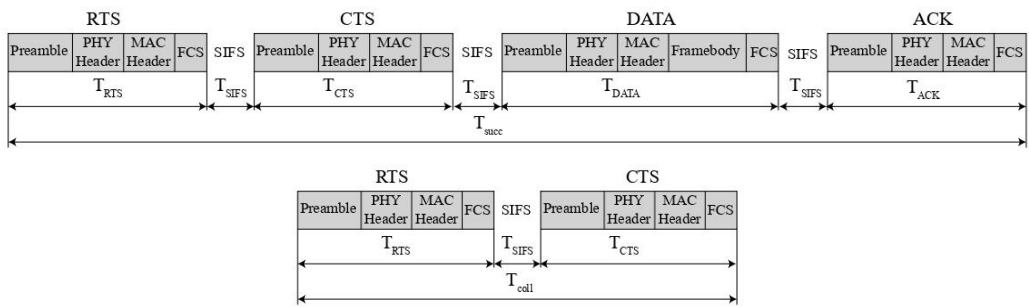

Figure 4. Success time and collision time of RTS/CTS access mechanism for the modified IEEE 802.15.6 MAC scheme

Table 1. Comparison of MAC parameters

| MAC parameters | Standard IEEE 802.15.6 MAC scheme | | | | | | | | Modified IEEE 802.15.6 MAC scheme | | | |
|---|---|---|---|---|---|---|---|---|---|---|---|---|
| $UP_i$ | $UP_0$ | $UP_1$ | $UP_2$ | $UP_3$ | $UP_4$ | $UP_5$ | $UP_6$ | $UP_7$ | $UP_0$ | $UP_2$ | $UP_4$ | $UP_6$ |
| $CW_{i,min}$ | 16 | 16 | 8 | 8 | 4 | 4 | 2 | 1 | 16 | 8 | 4 | 2 |
| $CW_{i,max}$ | 64 | 32 | 32 | 16 | 16 | 8 | 8 | 4 | 64 | 32 | 16 | 8 |
| Traffic designation | Background | Best Effort | Excellent Effort | Control Load | Video | Voice | High priority medical data or network control | Emergency or medical important report | Background and Best Effort | Excellent Effort and Control Load | Video and Voice | High priority medical data or network control and emergency or medical important report |
| Phases | EAP1, RAP1, MAP1, EAP2, RAP2, MAP2, CAP | | | | | | | | RAP, MAP, CAP | | | |
| Access mechanism | Basic access mechanism | | | | | | | | RTS/CTS access mechanism | | | |

## 5. ANALYTICAL MODEL

In this section, we first present a system model of the WBAN where the proposed MAC scheme and the standard IEEE 802.15.6 MAC scheme are deployed independently and then we design a Markov chain-based analytical model for the proposed MAC scheme.The notations that are used in the analytical model are summarized in Table 2.

Table 2. Notation used in the model

| Notation | Definition | Notation | Definition |
|---|---|---|---|
| $UP_i$ | User priority $i, i \in \{0, 2, 4, 6\}$ | $n_i$ | Number of nodes of $UP_i$ |
| $CW_{i,min}$ | Minimum contention window of $UP_i$ node | $T_{succ}$ | Duration of successful transmission |
| $CW_{i,max}$ | Maximum contention window of $UP_i$ node | $W_{i,j}$ | Contention window size of $UP_i$ at backoff stage $j$ |
| $T_{coll}$ | Duration of collision transmission | $S_{agg}$ | Aggregate throughput of network |
| $m_i$ | Maximum backoff stage of $UP_i$ beyond which the contention window will not be increased | $(i, j, k)$ | State of $UP_i$ node at Markov chain where backoff stage number $j$ and backoff counter value $k$ |
| $m_i + x_i$ | Retry limit of $UP_i$ node | $\lambda_i$ | Packet arrival rate of $UP_i$ node |
| $P_{i,succ}$ | Probability of successful transmission of $UP_i$ node | $P_{i,fail}$ | Probability of failure transmission of $UP_i$ node |
| $b_{i,j,k}$ | Stationary probability of $UP_i$ node at $(i, j, k)$ state | $P_{i,acce}$ | Channel access probability of $UP_i$ node |
| $b_{i,empty}$ | Stationary probability of $UP_i$ node at $(i, empty)$ state | $P_{i,drop}$ | Probability of packet drop of $UP_i$ node |
| $P_{i,idle}$ | Probability that a CSMA slot is idle sensed by a $UP_i$ node | $P_{idle}$ | Probability that a CSMA slot is idle |
| $P_{i,coll}$ | Probability of collision transmission of $UP_i$ node | $\rho_i$ | Probability that a $UP_i$ node has empty queue |
| $P_{i,error}$ | Probability of error transmission of $UP_i$ node | $\tau_i$ | Transmission probability of $UP_i$ node |
| $P_{tran}$ | Transmission probability | $T_{i,delay}$ | Average access delay of $UP_i$ node |
| $T_e$ | Expected time spent at each state | $E_i$ | Energy consumption of $UP_i$ node |
| $T_{csmaslot}$ | Length of CSMA slot in time | $T_{rap}$ | Length of RAP in time |
| $BER$ | Bit error rate | $S_i$ | Throughput of $UP_i$ node |
| $PER$ | Packet error rate | $T_{error}$ | Duration of error transmission |
| $F_J$ | Jain's fairness index | $U_i$ | Channel utilization of $UP_i$ node |
| $R_i$ | Reliability of $UP_i$ node | $U$ | Total channel utilization |
| $n_{UP}$ | Number of user priorities | $n$ | Total number of nodes |

### 5.1. System Model

In this paper, we consider a continuous healthcare monitoring system where a patient is equipped with one and only one central hub as coordinator and up to $n$ ($n = mMaxBANSize = 64$) number of identical medical sensor nodes, deployed on the body, which together forms a one-hop star topology of intra-WBAN, shown in Figure5. We mainly focus on the uplink frame transmission (from node to coordinator). The network consists of all four UPs ($UP_0$, $UP_2$, $UP_4$ and $UP_6$) traffic, where 0 denotes the lowest priority and 6 denotes highest priority traffic. It is considered that every node has a single queue that contains only one user priority data frame. The packet arrival process of all UPs is Poisson process with rate $\lambda$and this arrival rate is equal for all UPs. We assume that the coordinator operates in beacon mode withsuperframe boundaries where MAP and CAP are set to zero. It is also considered that all nodes and coordinator follow immediate acknowledgement (I-ACK) policy and use automatic repeat request (ARQ) as an error control method. A failure transmission occurs due to two reasons. The first reason is collision transmission, more than one nodes transmit at the same time and the second one is error transmission that occurs due to a noisy channel. A frame is dropped for all UPs when the number of failure transmission exceeds the finite retry limits ($m_i + x_i$). A node that has a data frame to transmit cannot generate a new data frame until either it receives the ACK frame for the transmitted data frame or the transmitted data frame is dropped. We assume that all UPs nodes have equal traffic load and payload size. We consider that collision probability of a frame that transmitted by a node is independent of the number of retries, i.e.,backoff stages. We also assume that a node transmits just one data frame after successfully access the channel. We use narrowband (NB) of IEEE 802.15.6 standard as a PHY layer to evaluate the performance of modified IEEE 802.15.6 MAC scheme and standard IEEE 802.15.6 MAC scheme. In this work, we ignore the hidden terminal, expose terminal and channel capture effect.

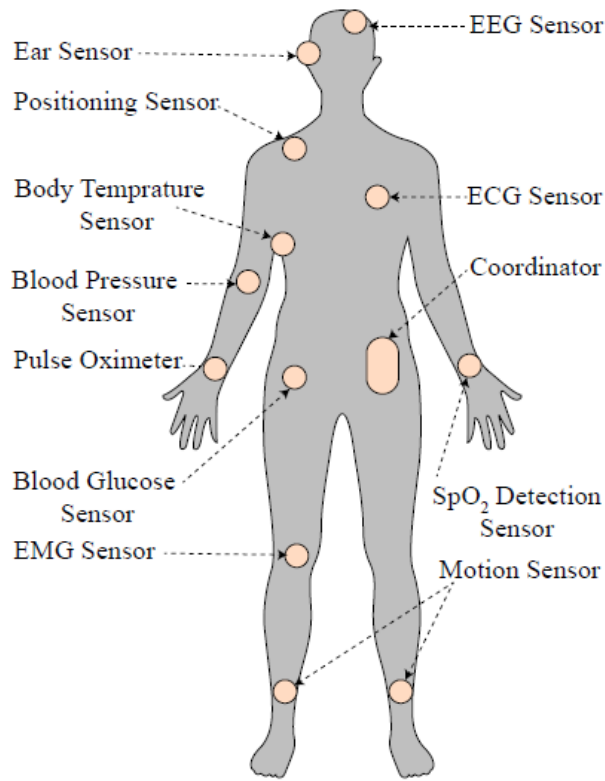

Figure 5. System model

## 5.2. Markov Chain Model

We develop a 2-D Markov chain model to describe the backoff procedure of modified IEEE 802.15.6 MAC scheme according to $M/G/1$ queuing model, which is shown in Figure 6. There are some studies where Markov chain-based analytical model were presented to evaluate the performance of CSMA/CA mechanism [8]–[18].In this Markov chain, the state of each$UP_i$ nodeis denoted by $(i, j, k)$where $i$,$j$and $k$ indicates user priority of the node, backoff stage numberand backoff counter value, respectively. The initial value of $j$for a new frame is 0 and is incremented by one after every failure transmission until it reaches the retry limit ($m_i + x_i$). After every successful transmission or frame drop, the value of $j$will be reset to 0.The value of $k$ is initially set with a value that is randomly chosen from $[1, W_{i,j}]$, where $W_{i,j}$ denotes the contention window size of a $UP_i$ node at backoff stage $j$. The backoff counter value $k$ is decremented by one if the channel is sensed idle in a CSMA slot and if there is sufficient time to complete current frame transmission before the end of the current access phase. When the counter value becomes zero, frame is transmitted immediately.If failure transmission is occurred due to transmission collision or error transmission, the node goes to the next backoff stage with a new backoff counter value. If frame is successfully transmitted, the node select a new backoff counter value under initial backoff stage. The contention window size of $UP_i$ node at backoff stage$j$ is expressed as

$$W_{i,j} = \begin{cases} CW_{i,min} & ; \quad \boldsymbol{j = 0} \\ 2^{j/2} \times CW_{i,min} & ; \quad \boldsymbol{1 \leq j \leq m_i} \text{ and } \boldsymbol{j} \text{ is even} \\ 2^{(j-1)/2} \times CW_{i,min} & ; \quad \boldsymbol{1 \leq j \leq m_i} \text{ and}\boldsymbol{j} \text{ is odd} \\ CW_{i,max} & ; \quad \boldsymbol{m_i < j \leq m_i + x_i} \end{cases} \qquad (1)$$

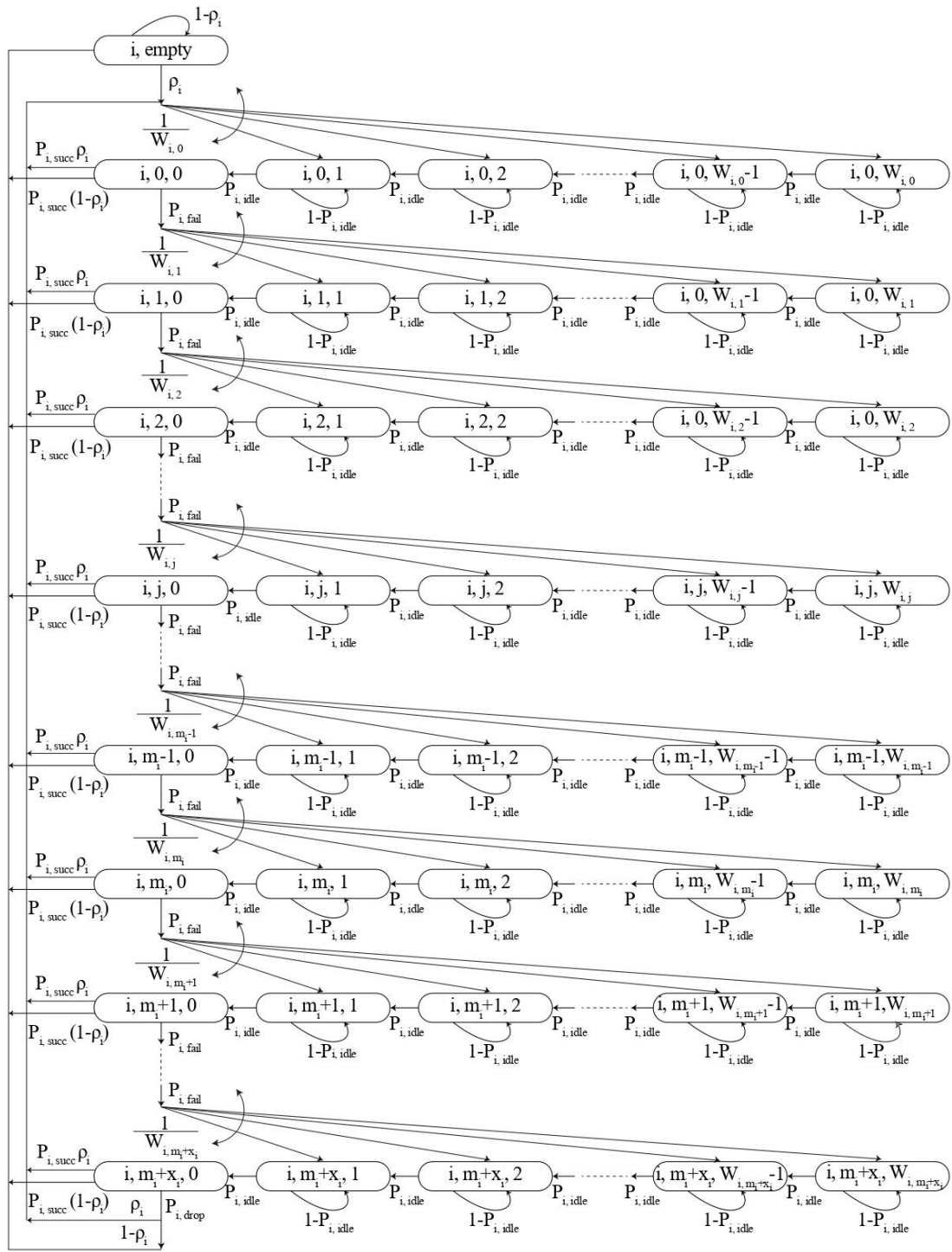

Figure 6. Markov chain for modified IEEE 802.15.6 MAC scheme of $UP_i$ node in saturated and non-saturated traffic conditions

To analyse both saturation and non-saturation traffic conditions of node, we introduce a state in the Markov chain is denoted by $(i, empty)$, which represents the state of the $UP_i$ node when the node is empty after a successful transmission or a frame drop. We also define$b_{i,j,k}$thatrepresents the stationary probabilities of states $(i, j, k)$. Now we derive the stationary probabilities $b_{i,j,k}$and $b_{i,empty}$ in terms of $b_{i,0,0}$ state. According to the Markov chain, the general expression of thestationary probabilities for zero backoff stage can be written in terms of $b_{i,0,0}$state, i.e.,

$$b_{i,0,k} = \frac{W_{i,0} - k + 1}{W_{i,0}} \frac{1}{P_{i,idle}} b_{i,0,0} \qquad ; \qquad \boldsymbol{1 \leq k \leq W_{i,0}} \qquad (2)$$

Again, the general expression of stationary probabilities for $j$th backoff stage can be writtenin terms of$b_{i,0,0}$ state,i.e.,

$$b_{i,j,k} = \frac{W_{i,j} - k + 1}{W_{i,j}} \frac{P_{i,fail}}{P_{i,idle}} b_{i,j-1,0} \qquad ; \qquad \boldsymbol{1 \leq j \leq m_i + x_i} \text{and } \boldsymbol{1 \leq k \leq W_{i,j}} \qquad (3)$$

The last state of each backoff stage can be also writtenin terms of$b_{i,0,0}$state, i.e.,

$$b_{i,j,0} = (P_{i,fail})^j b_{i,0,0} \qquad ; \qquad \boldsymbol{1 \leq j \leq m_i + x_i} \qquad (4)$$

By substituting Eq. (2) and Eq.(4) in Eq. (3), the general expression of the stationary probability for each backoff stage of the Markov chain can be obtained in terms of $b_{i,0,0}$ state, i.e.,

$$b_{i,j,k} = \frac{W_{i,j} - k + 1}{W_{i,j}} \frac{(P_{i,fail})^j}{P_{i,idle}} b_{i,0,0} \qquad ; \qquad \boldsymbol{0 \leq j \leq m_i + x_i} \text{ and } \boldsymbol{1 \leq k \leq W_{i,j}} \qquad (5)$$

We define $\rho_i$ as the probability that the node queue of $UP_i$has at least one frame waiting for transmission. The stationary probability of empty state can be obtained from the Markov chain as

$$b_{i,empty} = (1 - \rho_i) b_{i,empty} + \sum_{j=0}^{m_i+x_i-1} (1 - \rho_i) P_{i,succ} b_{i,j,0} + (1 - \rho_i) b_{i,m_i+x_i,0} \qquad (6)$$

By substituting Eq. (5) in Eq. (6), the stationary probability of empty state can be obtained in terms of $b_{i,0,0}$ state, i.e.,

$$b_{i,empty} = \frac{(1-\rho_i)}{\rho_i} b_{i,0,0} \tag{7}$$

Now, the normalization condition of the Markov chain can be expressed as

$$\sum_{j=0}^{m_i+x_i} b_{i,j,0} + \sum_{k=1}^{W_{i,j}} b_{i,0,k} + \sum_{j=1}^{m_i+x_i} \sum_{k=1}^{W_{i,j}} b_{i,j,k} + b_{i,empty} = 1 \tag{8}$$

By substituting Eq. (2)–Eq. (4) and Eq. (7) in Eq. (8), the stationary probability of $(i, 0, 0)$ state can be obtained, i.e.,

$$b_{i,0,0} = \frac{1}{\frac{(1-P_{i,fail})^{m_i+x_i+1}}{1-P_{i,fail}} + \sum_{j=0}^{m_i+x_i} \frac{W_{i,j}+1}{2} \frac{(P_{i,fail})^j}{P_{i,idle}} + \frac{(1-\rho_i)}{\rho_i}} \tag{9}$$

We assume that the retry limit of CSMA/CA mechanism is finite. When number of failure transmission of $UP_i$ node exceed the retry limit $(m_i + x_i)$, the frame is dropped and the node initiates the backoff procedure again to transmit a new frame. Thus, the frame drop probability is written as

$$P_{i,drop} = (P_{i,fail})^{m_i+x_i+1} \tag{10}$$

According to the CSMA/CA mechanism, a node will attempt for transmission when the node is in $(i, j, 0)$, $0 \le j \le m_i + x_i$, state of the Markov chain. Thus the transmission probability of each $UP_i$ node is written as

$$\tau_i = \sum_{j=0}^{m_i+x_i} b_{i,j,0} \tag{11}$$

By substituting Eq. (4) in Eq. (11), the transmission probability can be obtained in terms of $b_{i,0,0}$ state, i.e.,

$$\tau_i = \frac{(1-P_{i,fail})^{m_i+x_i+1}}{1-P_{i,fail}} b_{i,0,0} \tag{12}$$

According to the CSMA/CA mechanism of IEEE 802.15.6 standard, we define $P_{i,lock}$ as the probability that in a given CSMA slot if there is not enough time for transmitting a frame in the current permitted access period. At this time the backoff counter shall be freeze till the beginning of the next permitted access period. Thus, the counter lock probability is estimated as

$$P_{i,lock} = \frac{1}{L_{rap} - L_{succ} - C_i} \tag{13}$$

Where $L_{rap}$ and $L_{succ}$ denote the length of RAP in slots, the duration of successful transmission time in slots, respectively. The mean backoff value of $UP_i$ node $(C_i)$ is approximated as

$$C_i = \frac{\frac{m_i}{2} + W_{i,j}\left(2^{\frac{m_i}{2}} - 1\right) + (x_i + 1)\frac{W_{i,j}2^{\frac{m_i}{2}+1}}{2}}{m_i + x_i + 1} \tag{14}$$

The transmission probability in a CSMA slotis that at least one node transmits a frame through themedium. Thus, we have

$$P_{tran} = 1 - (1 - \tau_0)^{n_0}(1 - \tau_2)^{n_2}(1 - \tau_4)^{n_4}(1 - \tau_6)^{n_6} \tag{15}$$

Where $n_i$ denotes the number of nodes of $UP_i$ in the network. We assume that number of nodes of each user priority is equal,$n_i = n/n_{UP}$, where $n$ is the total number of nodes in the network and $n_{IIP}$ is the number of user priorities. Theidle probability of a CSMA slotis that no node is transmitting in the slot. Thus, we have

$$P_{idle} = (1 - \tau_0)^{n_0}(1 - \tau_2)^{n_2}(1 - \tau_4)^{n_4}(1 - \tau_6)^{n_6} \tag{16}$$

The idle probability of a CSMA slotsensed by a $UP_i$ node is that the medium remains idle in a CSMA slot during the backoff procedure of the $UP_i$ node and the backoff counter is unlock. In another word, this is the probability of counter down of $UP_i$ node. Thus we have,

$$P_{i,idle} = \frac{(1 - \tau_0)^{n_0}(1 - \tau_2)^{n_2}(1 - \tau_4)^{n_4}(1 - \tau_6)^{n_6}}{(1 - \tau_i)}(1 - P_{i.lock}) \tag{17}$$

There is the significant difference between channel access probability and successful transmission probability. The channel access probability of $UP_i$ node is the probability that the channel is successfully accessed by the$UP_i$ node, conditioned on the fact that other nodes are not transmitting. Thus, we have

$$P_{i,acce} = \frac{n_i\tau_i(1 - \tau_0)^{n_0}(1 - \tau_2)^{n_2}(1 - \tau_4)^{n_4}(1 - \tau_6)^{n_6}}{(1 - \tau_i)(1 - P_{idle,rap})} \tag{18}$$

The successful transmission probability of $UP_i$ nodeis the probability that a$UP_i$ node access the channel successfully and receive ACK from the coordinator. Thus, we have

$$P_{i,succ} = P_{i,acce}(1 - PER) \tag{19}$$

Where $PER$ indicates the packet error rate whose value depends on access mechanism of IEEE 802.15.6 standard. We assume that immediate acknowledgement (I-ACK) is used in the network and a node transmits just one data frame during channel access period and thus, the packet error rate for the RTS/CTS access mechanismis written as

$$PER = 1 - (1 - BER)^{L_{RTS}+L_{CTS}+L_{DATA}+L_{ACK}} \tag{20}$$

Where $BER$ denotes the bit error rate of the channel and$L_{RTS}$, $L_{CTS}$, $L_{DATA}$ and $L_{ACK}$ are the lengths of RTS, CTS, DATA and ACK frame in bits. After successfully access the channel, the transmission of a node is failed due to two reasons. The first reason is collision in transmission and the second one is error in transmission due to noisy channel. Therefore, the transmission failure probability of $UP_i$node is written as

$$P_{i,fail} = P_{i,coll} + P_{i,error} \tag{21}$$

where $P_{i,coll}$ and $P_{i,error}$ represent the transmission collision probability and transmission error probability in RAP, respectively. The transmission collision occurs when at least two nodes transmit frame at the same time over the channel. The transmission collision probability of $UP_i$ node is written as

$$P_{i,coll} = 1 - \frac{n_i \tau_i (1-\tau_0)^{n_0}(1-\tau_2)^{n_2}(1-\tau_4)^{n_4}(1-\tau_6)^{n_6}}{(1-\tau_i)(1-P_{idle})} = 1 - P_{i,acce} \qquad (22)$$

The transmission error occurs when the channel is accessed successfully as well as the frame is transmitted but the frame contains erroneous bit. Thus, we have

$$P_{i,error} = P_{i,acce} PER \qquad (23)$$

By substituting Eq. (22) and Eq.(23) in Eq. (21), the failure transmission probability can be obtained as

$$P_{i,fail} = 1 - P_{i,acce}(1 - PER) \qquad (24)$$

In this work, since we assumed that packet arrival rate of $UP_i$ node is Poisson process with rate $\lambda_i$, the probability that the node has at least one packet waiting for transmission can be determined as

$$\rho_i = 1 - e^{-\lambda_i T_e} \qquad (25)$$

Where $T_e$ represents the expected time spent by node at each state. The duration of this time is not fixed and depends on the channel state, transmission state and access phase. If the channel is idle, the duration of the state is one CSMA slot $T_{csmaslot}$. When the channel is sensed as busy its means that either successful transmission, transmission collision, or error transmission is occurred in the channel. If successful transmission is occurred, the duration of the state is the time of a successful transmission, which is estimated as

$$T_e = T_{csmaslot}(1 - P_{tran}) + T_{succ}P_{tran}P_{succ} + T_{coll}P_{tran}P_{coll} + T_{error}P_{tran}P_{error} \qquad (26)$$

Where $P_{succ}$, $P_{coll}$ and $P_{error}$ are the total successful transmission probability, total transmission collision probability and total error transmission probability of the node in the network, respectively. Thus, we have

$$P_{succ} = P_{0,succ} + P_{2,succ} + P_{4,succ} + P_{6,succ} \qquad (27)$$
$$P_{coll} = P_{0,coll} + P_{2,coll} + P_{4,coll} + P_{6,coll} \qquad (28)$$
$$P_{error} = P_{0,error} + P_{2,error} + P_{4,error} + P_{6,error} \qquad (29)$$

According to RTS/CTS access mechanism, duration of successful transmission ($T_{succ}$), collision transmission ($T_{coll}$) and error transmission ($T_{error}$) are expressed as

$$T_{coll} = T_{RTS} + T_{CTS} + T_{SIFS} + 2\alpha \qquad (30)$$

It is observed that for RTS/CTS access mechanism, detection of transmission error or successful transmission require same amount of time in Figure 4. Thus, we have

$$T_{error} = T_{succ} = T_{RTS} + T_{CTS} + T_{DATA} + T_{ACK} + 3T_{SIFS} + 4\alpha \qquad (31)$$

where $\alpha$ denotes the propagation delay and $T_{RTS}$, $T_{CTS}$, $T_{DATA}$, $T_{ACK}$ and $T_{SIFS}$ represent the duration of RTS, CTS, DATA, ACK and SIFS in time, respectively and that are determined as

$$T_{DATA} = \frac{Preamble}{R_S} + \frac{PHY\ Header}{R_{PLCP}} + \frac{(MAC\ Header + Framebody + FCS) \times 8}{R_{PSDU}} \quad (32)$$

$$T_{RTS/CTS/ACK} = \frac{Preamble}{R_S} + \frac{PHY\ Header}{R_{PLCP}} + \frac{(MAC\ Header + FCS) \times 8}{R_{PSDU}} \quad (33)$$

Simplifying all the above derived equations of the proposed analytical model, weobtain 21 equations while we have 21 unknown variables:$P_{idle}$, $P_{i,idle}$, $P_{i,acce}$, $P_{i,fail}$, $\tau_i$ and $\rho_i$.Finally, we solve the simplified equations using Maple [22] and then determine the performance metrics that are defined in following Section.

# 6. PERFORMANCE METRICS

In this section, we derive the expressions of the performance metrics viz. reliability, throughput, aggregate throughput, average access delay, energy consumption, channel utilization and fairness to explore the performance of the network.

## 6.1. Reliability

The reliability ($R_i$) of $UP_i$node is defined as the complementary probability with which a transmitted packet is dropped due to repeated failure transmission after $m_i + x_i + 1$ attempts [8]. Therefore, $R_i$ is mathematically expressed as

$$R_i = 1 - P_{i,drop} = 1 - \left(P_{i,fail}\right)^{m_i + x_i + 1} \quad (34)$$

## 6.2. Throughput

The throughput ($S_i$) of $UP_i$nodeis defined as the average number of bits successfully transmitted by the $UP_i$ node persecond [23]–[24]. Therefore, $S_i$ is mathematically expressed as

$$S_i = \frac{P_{i,succ}P_{tran}L_{framebody}}{T_e} \quad (35)$$

## 6.3. Aggregated Throughput

The aggregated throughput ($S_{agg}$) of the network is the summation of throughput of all$UP_i$ nodes of the network [25]. Therefore, $S_{agg}$is expressed as

$$S_{agg} = S_0 + S_2 + S_4 + S_6 \quad (36)$$

## 6.4. Average Access Delay

The average access delay ($T_{i,delay}$) of $UP_i$ node is defined as the average time from the instant when frame is generated to the instant when the packet is successfully transmitted or dropped [26]. Therefore, $T_{i,delay}$ is mathematically expressed as

$$T_{i,delay} = \sum_{j=0}^{m_i+x_i} \left(P_{i,fail}\right)^j \left(1 - P_{i,fail}\right) \sum_{l=0}^{j} \frac{W_{i,l}+1}{2} T_e + \left(P_{i,fail}\right)^{m_i+x_i+1}$$
$$\sum_{l=0}^{m_i+x_i+1} \frac{W_{i,l}+1}{2} T_e \tag{37}$$

## 6.5. Energy Consumption

The energy consumption of a node is affected by several stages: idle stage, successful stage, collision stage and error stage [27]– [29]. Assume that $P_{TX}$, $P_{RX}$ and $P_{IDLE}$ denote the energy consumption in transmitting state, receiving state and idle state of a node, respectively. The mean energy consumption ($E_i$) of $UP_i$ node is expressed as

$$E_i = E_{i,idle} + E_{i,succ} + E_{i,coll} + E_{i,error} \tag{38}$$

The each components of mean energy consumption are defined in Eq. (39) – Eq.(42).

$$E_{i,idle} = T_{csmaslot} P_{IDLE} P_{i,idle} \tag{39}$$

$$E_{i,succ} = (T_{RTS}P_{TX} + T_{CTS}P_{RX} + T_{DATA}P_{TX} + T_{ACK}P_{RX} + 3T_{SIFS}P_{IDLE})P_{tran}P_{i,succ}$$
$$+ T_{succ}P_{IDLE}P_{tran}(P_{succ} - P_{i,succ}) \tag{40}$$

$$E_{i,coll} = (T_{RTS}P_{TX} + T_{CTS}P_{RX} + T_{SIFS}P_{IDLE})P_{tran}P_{i,coll} + T_{coll}P_{IDLE}P_{tran}$$
$$(P_{coll} - P_{i,coll}) \tag{41}$$

$$E_{i,error} = (T_{RTS}P_{TX} + T_{CTS}P_{RX} + T_{DATA}P_{TX} + T_{ACK}P_{RX} + 3T_{SIFS}P_{IDLE})P_{tran}$$
$$P_{i,error} + T_{succ}P_{IDLE}P_{tran}(P_{error} - P_{i,error}) \tag{42}$$

## 6.6. Channel Utilization

The channel utilization ($U_i$) of $UP_i$ node is defined as the ratio of the time that the $UP_i$ node uses for successful transmission of data frame at a state and the time that the $UP_i$ node stays at that state [25]. The channel utilization ($U_i$) of $UP_i$ node is expressed as

$$U_i = \frac{P_{i,succ}P_{tran}T_{framebody}}{T_e} \tag{43}$$

The overall channel utilization ($U$) is the summation of channel utilization of all $UP_i$ nodes. Therefore, $U$ is expressed as

$$U = U_0 + U_2 + U_4 + U_6 \tag{44}$$

## 6.7. Fairness

The Jain's fairness index [30] is used to measure the fairness in resource allocation among different UPs nodes and it mathematically expressed as

$$F_J = \frac{(S_0 + S_2 + S_4 + S_6)^2}{N_{UP}(S_0{}^2 + S_2{}^2 + S_4{}^2 + S_6{}^2)} \tag{45}$$

# 7. PERFORMANCE EVALUATION

We consider a general scenario of intra-WBAN to evaluate the performance metrics viz.reliability, throughput, aggregate throughput, average access delay, energy consumption, channel utilization and fairness. The network composes of n numbers of identical medical sensors and a coordinator that use narrowband (NB) as a PHY layer. We conduct two experiments to compare the performance of standard IEEE 802.15.6 MAC scheme and modified IEEE 802.15.6 MAC scheme.One experiment uses the standard IEEE 802.15.6 MAC scheme as a MAC layer protocol and another uses the modified IEEE 802.15.6 MAC scheme as a MAC layer protocol.In the first experiment, we set the access phases length of the superframe for standard IEEE 802.15.6 MAC scheme as EAP1 = 0.1 sec and RAP1= 0.8 secwhile MAP1, EAP2, RAP2, MAP2 and CAP areset to zero. In the second experiment, we set the access phases length of the superframe for modified IEEE 802.15.6 MAC scheme as RAP = 0.9 sec whereas MAP and CAP are set to zero.

Table 3. System parameters

| Parameters | Values | Parameters | Values | Parameters | Values |
|---|---|---|---|---|---|
| Preamble | 90 bits | $R_{PLCP}$ | 91.9 kbps | Retry limit | 7 |
| PHY Header | 31 bits | $R_{PSDU}$ | 971.4 kbps | $BER$ | $2 \times 10^{-5}$ |
| MAC Header | 56 bits | $R_S$ | 600 kbps | Frequency band | 2.4 GHz |
| FCS | 16 bits | $T_{RAP}$ | 1 s | Modulation | DBPSK |
| RTS | 193 bits | $T_{csmaslot}$ | 125 µs | $\lambda_i$ | 0.5 pkts/sec |
| CTS | 193 bits | SIFS | 75 µs | $P_{TX}$ | 27 mW |
| ACK | 193 bits | α | 1 µs | $P_{RX}$ | 1.8 mW |
| *Framebody* | 800 bits | $N_{UP}$ | 4 | $P_{IDLE}$ | 5 µW |

Figure 7 compares the reliability per UP of standard IEEE 802.15.6 MAC scheme and modified IEEE 802.15.6 MAC scheme where reliability is instructed as the function of number of nodes.It is seen that reliability per UP decreases as the number of nodes increases for both MAC scheme. With the number of nodes increasing, more nodes will contend for transmission, which would result in more collision and thus a reduction in reliability.It is also observed that reliability is highest in$UP_7$for standard IEEE 802.15.6 MAC scheme whereas$UP_6$achieve highest reliability for modified IEEE 802.15.6 MAC scheme and reliability decreases with the decrease of the user priority for both MAC scheme.In MAC layer the user priorities are differentiated by the value of minimum and maximum contention window size. The contention window size gradually increases from $UP_7$ to $UP_0$, shown in Table 1. Due to larger value of contention window of lower user priorities, lower user priorities nodes get less chance to deliver frames compared to higher user priorities nodes and thus a reduction in reliability. The results clearly show that the modified IEEE 802.15.6 MAC scheme provides 90% reliability whereas below 70% reliability is achieved except highest user priority $UP_7$ when the network has smallnumber of nodes. This is because the $UP_7$ nodes only access the EAP phase in standard IEEE 802.15.6 MAC scheme. Due to the larger contention window size of proposed user priorities and use of RTS/CTS access mechanism, the modified IEEE 802.15.6 MAC scheme provides better reliability than the standard IEEE 802.15.6 MAC scheme when the network is also insaturated state.

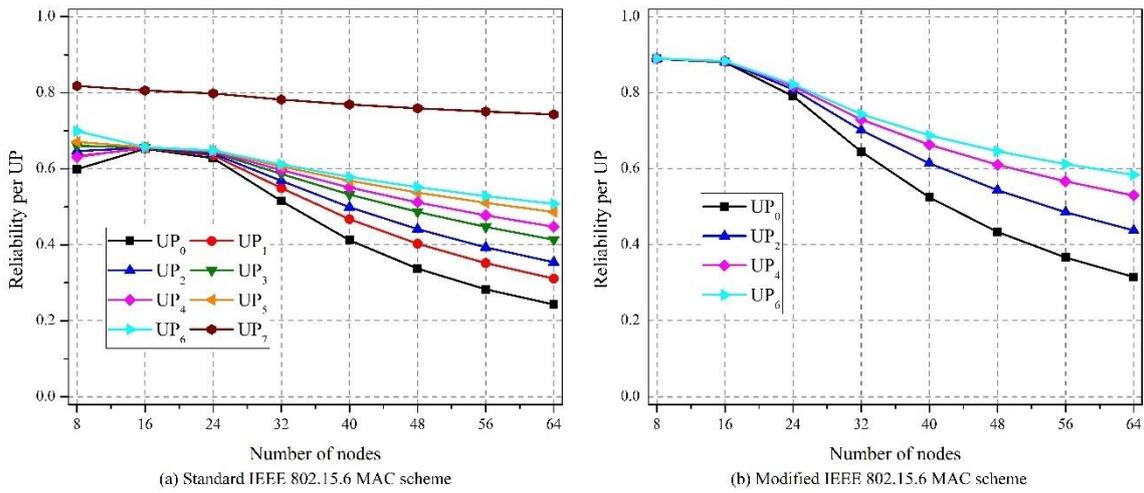

(a) Standard IEEE 802.15.6 MAC scheme          (b) Modified IEEE 802.15.6 MAC scheme

Figure 7. Comparing reliability per UP of standard IEEE 802.15.6 MAC scheme and modified IEEE 802.15.6 MAC scheme

Figure 8 compares the throughput per UP of standard IEEE 802.15.6 MAC scheme and modified IEEE 802.15.6 MAC scheme where throughput is instructed as the function of number of nodes.It is observed that the throughput is highest in $UP_7$ forstandard IEEE 802.15.6 MAC scheme. This is because highest user priority nodes content for transmitting frame both EAP1 and RAP1 access phase with lowest contention window size.It is also noted thatthe throughput decreases with the decrease of the user priority in both scheme.Due to larger value of contention window of lower user priorities, lower user priorities nodes achieve less chance to transmit frame compared to higher user priorities nodes and thus a decrease in throughput.Another important observation is that modified IEEE 802.15.6 MAC scheme achieves more throughput compared to standard IEEE 802.15.6 MAC scheme for the same number of nodes in the network. This is because the proposed user priorities nodes contain larger contention window size and use the RTS/CTS access mechanism.

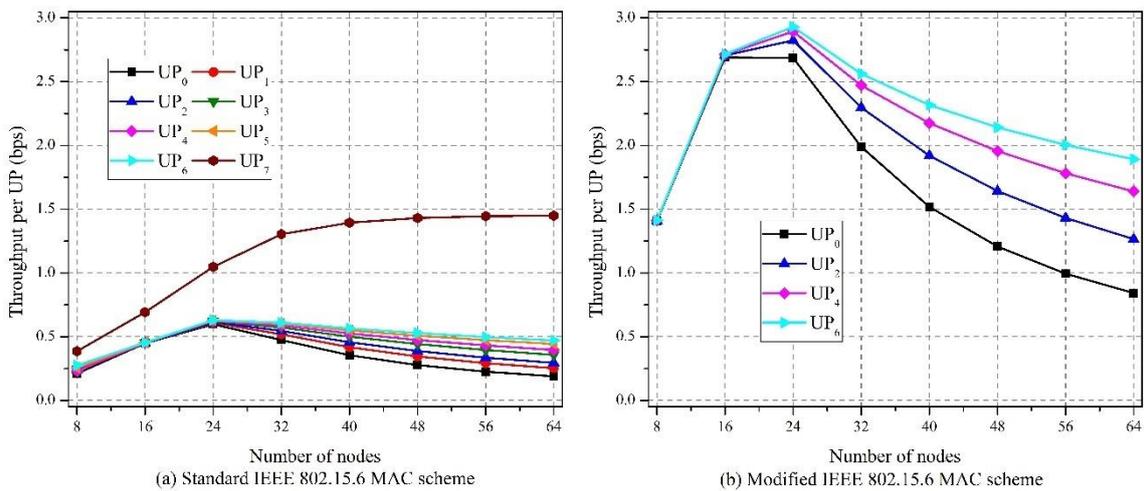

(a) Standard IEEE 802.15.6 MAC scheme          (b) Modified IEEE 802.15.6 MAC scheme

Figure 8. Comparing throughput per UP of standard IEEE 802.15.6 MAC scheme and modified IEEE 802.15.6 MAC scheme

Figure 9 compares the average access delay of standard IEEE 802.15.6 MAC scheme and modified IEEE 802.15.6 MAC scheme where average access delay is instructed as the function of number of nodes. It is observed that the average access delay is highest for $UP_0$ and it decreases with the increase of the user priority for both MAC scheme.Due to larger value of contention window of lower user priorities, lower user priorities nodes need more time to access the channel compared to higher user priorities nodes and thus an increase in average access delay. It is also seen that average access delay per UP is close to zero up to 16 nodes for both MAC scheme. This is because the small number of nodes leads to the network is in non-saturated state. With the number of nodes further increasing, more nodes will try to transmit frame simultaneously, which would cause more collisions and thus an exponential increase of average access delay. Another important observation is that the modified IEEE 802.15.6 MAC scheme offer less average access delay compared to standard IEEE 802.15.6 MAC scheme for same number of nodes in the network because of less collision probability of RTS/CTS access mechanism used in proposed MAC scheme.Figure 10 compares the energy consumption per UP of standard IEEE 802.15.6 MAC scheme and modified IEEE 802.15.6 MAC scheme where energy consumption is instructed as the function of number of nodes. It is observed that the energy consumption is same for $UP_0$to$UP_6$and lowest for $UP_7$in standard IEEE 802.15.6 MAC scheme. This is because $UP_0$to$UP_6$nodes are assigned to only RAP1 access phase and collision probability in RAP1 phase are almost same. The highest user priority nodes are assigned to both EAP1 and RAP1 access phase and suffer less collision compared to other user priority nodes. It is also seen that energy consumption exponentially increases as the number of nodes increases in the network for both MAC scheme. Due to increase of nodes in the network, more collision are occurred and thus increases of energy consumption. It is visualized that modified IEEE 802.15.6 MAC scheme consumes less energy compared to standard IEEE 802.15.6 MAC scheme for same number of nodes in the network due to the use of RTS/CTS access mechanism.Figure 11 compares the channel utilization per UP of standard IEEE 802.15.6 MAC scheme and modified IEEE 802.15.6 MAC scheme where channel utilization is instructed as the function of number of nodes.It is seen that the channel utilization is highest in $UP_7$ for standard IEEE 802.15.6 MAC scheme and it decreases with the decrease of the user priority.This is because highest user priority nodes content for transmitting frame both EAP1 and RAP1 access phase with lowest contention window size. Due to larger value of contention window of lower user priorities, lower user priorities nodes achieve less opportunity to transmit frame compared tohigher userpriorities nodes and thus a decrease in channel utilization.It is also observed that the modified IEEE 802.15.6 MAC scheme gains more channel utilization compared to standard IEEE 802.15.6 MAC scheme for same number of nodes in the network. Figure 12 compares the total channel utilization, Jain's fairness index and aggregate throughput of standard IEEE 802.15.6 MAC scheme and modified IEEE 802.15.6 MAC scheme where each performance metricis instructed as the function of number of nodes.Due to presence of EAP1 phase and the lowest contention window size, the highest user priority $UP_7$ nodes aggressively achieve highest throughputcompared to other UPs and thus fairness decreases in standard IEEE 802.15.6 MAC scheme.The modified IEEE 802.15.6 MAC scheme offers higher total channel utilization and aggregate throughput compared to standard IEEE 802.15.6 MAC scheme. This is becausethe modified IEEE 802.15.6 MAC scheme uses RTS/CTS access mechanism whereas basic access mechanism uses standard IEEE 802.15.6 MAC scheme. It is summarized that the modified IEEE 802.15.6 MAC scheme improves the total channel utilization, Jain's fairness index and aggregate throughput as compared to standard IEEE 802.15.6 MAC scheme.

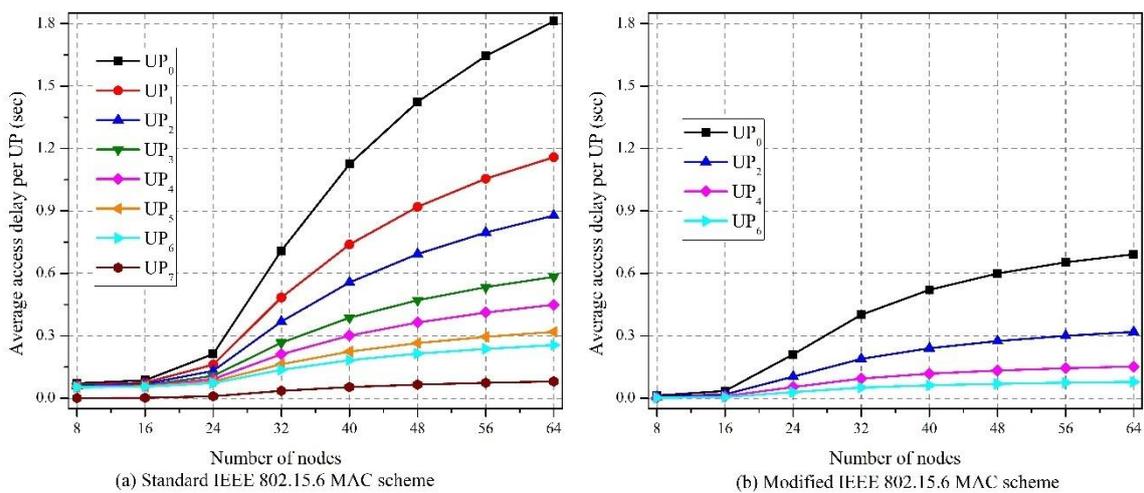

(a) Standard IEEE 802.15.6 MAC scheme      (b) Modified IEEE 802.15.6 MAC scheme

Figure 9. Comparing average access delay per UP of standard IEEE 802.15.6 MAC scheme and modified IEEE 802.15.6 MAC scheme

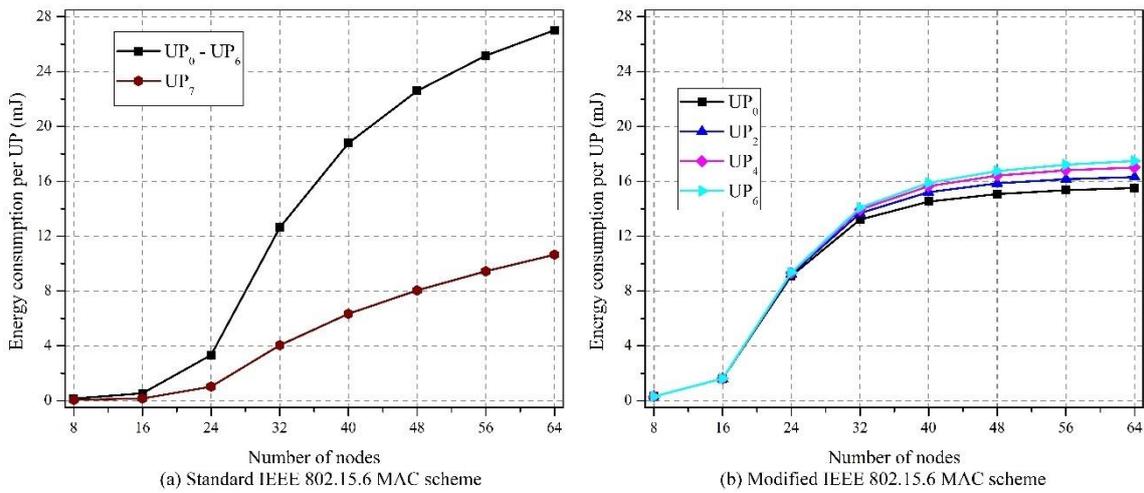

Figure 10. Comparing energy consumption per UP of standard IEEE 802.15.6 MAC scheme and modified IEEE 802.15.6 MAC scheme

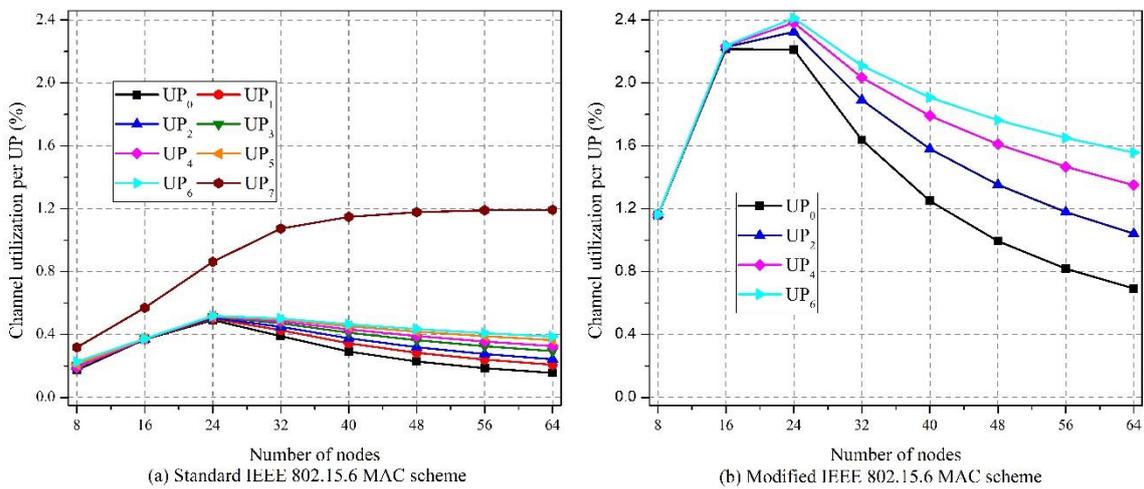

Figure 11. Comparing channel utilization per UP of standard IEEE 802.15.6 MAC scheme and modified IEEE 802.15.6 MAC scheme

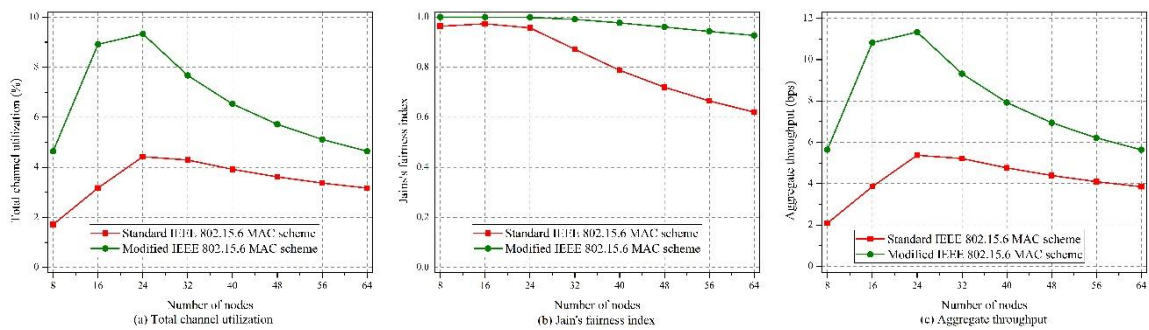

Figure 12. Comparing total channel utilization, Jain's fairness index and aggregate throughput of standard IEEE 802.15.6 MAC scheme and modified IEEE 802.15.6 MAC scheme

## 8. CONCLUSIONS

In this paper, we proposed a modified IEEE 802.15.6 MAC scheme in which the superframe structure is improved by excluding redundant and occasionally used access phases:EAP1, EAP2, RAP2 and MAP2.In the proposed MAC scheme, the eight user priorities of IEEE 802.15.6 standard are combined into four user priorities andRTS/CTS access mechanism are used instead of basic access mechanism.Moreover, we designed an analytical model to evaluate the performance during RAP access phases and solved the analytical model by Maple.We compared the analytical results of the modified IEEE 802.15.6 MAC scheme with standard IEEE 802.15.6 MAC scheme. The analytical results clearly indicated thatthe modified IEEE 802.15.6 MAC scheme outperforms interms of reliability, throughput, access delay, energy consumption, utilization and fairness compared to standard IEEE 802.15.6 MAC scheme.The proposed MAC scheme could be effective in E-health applications of WBAN but it narrows the scope of WBAN applications.In future, an adaptive MAC scheme will be designed to improve performance by dynamically adjusting the length of access phases according to the traffic load.

## CONFLICTS OF INTEREST

The authors declare that they have no conflict of interest regarding the publication of this paper.

## ACKNOWLEDGEMENTS

This work was supported by the Institute of Research and Training (IRT), Hajee Mohammad Danesh Science and Technology University (HSTU), Dinajpur, Bangladesh.

## AUTHORS

**Md. Abubakar Siddik** received the B. Sc. from the department of Telecommunication and Electronic Engineering (currently ECE), Hajee Mohammad Danesh Science and Technology University (HSTU), Dinajpur-5200, Bangladesh in 2011. He also received the M. Sc. from Institute of Information and Communication Technology (IICT), Bangladesh University of Engineering and Technology (BUET), Dhaka-1205, Bangladesh in 2017. He worked as a Lecturer in Prime University, Dhaka, Bangladesh and City University, Dhaka, Bangladesh from 2013-2014. He joined as a Lecturer in department of Electronics and Communication Engineering (ECE), HSTU in 2014 and currently he is working as an Assistant Professor in the same department. His current research interests include communication protocol of IoT, Smart Grid, UAV, UWSN, WBAN, VANET, in particular, design and performance analysis of communication protocol via simulation and analytical model.
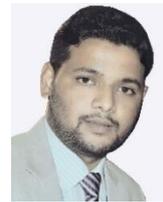

Most. **Anju Ara Hasi** received the B. Sc. (Engineering) from the department of Electronics and Communication Engineering (ECE), Hajee Mohammad Danesh Science and Technology University (HSTU), Dinajpur-5200, Bangladesh in 2017. Currently, she is a M. Sc. (Engineering) thesis semester student in the same department. Her research interest is performance analysis of communication protocols of IoT and WBAN.
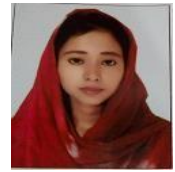

**Jakia Akter Nitu** received the B. Sc. (Engineering) from the department of Electronics and Communication Engineering (ECE), Hajee Mohammad Danesh Science and Technology University (HSTU), Dinajpur-5200, Bangladesh in 2019. Currently, she is a student of M. Sc. (Engineering) in the same department. Her research interest is performance analysis of communication protocols of IoT, UAV, WBAN and WSN.
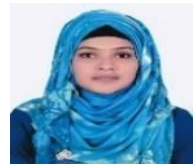

**Sumonto Sarker** has completed his B. Sc. from Hajee Mohammad Danesh Science and Technology University and Masters from University of Dhaka. He attends a meritorious career in both while student and working. Now he is working as Associate Professor in Hajee Mohammad Danesh Science and Technology University. He has developed several network related works, and has most interest in Under Water Target Tracking, design different types routing protocol about MANET, VANET.
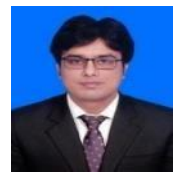

**Nasrin Sultana** received the B. Sc. from Hajee Mohammad Danesh Science and Technology University (HSTU), Dinajpur-5200, Bangladesh in 2009 and Masters from Bangladesh University of Engineering and Technology (BUET), Dhaka-1205, Bangladesh in 2014. She has been awarded the Doctor of Philosophy (Ph.D) degree from the Graduate school of Science and Engineering, Saitama University, Japan in 2020. Now she is working as an Associate Professor in the department of Electronics and Communication Engineering, HSTU, Dinajpur. Her research interest is in high-speed optical measurement and sensing in ultrafast photonic science.
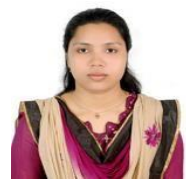

**Emran Ali** is an Assistant Professor at Hajee Mohammad Danesh Science and Technology University (HSTU), Bangladesh. He has been working as an academic staff at HSTU since 2014. He also has experience in software development and management for three years. He received his Bachelor of Science degree from the Department of Computer Science and Engineering at HSTU. He finished his Master of Science (by research) degree in Information Technology with a Biomedical Engineering 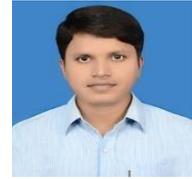 specialization from Deakin University, Australia. His main research areas are sensor and biosensor-based signal processing, particularly brain signal processing. Epileptic seizure detection, prediction, and Electroencephalogram channel optimization for epileptic seizure classification are his main works. His research also includes observing different (brain) disorders with sleep stages and sleep patterns.